\documentstyle[epsfig,preprint,aps]{revtex}
\begin{document}
\tighten

\def\bea{\begin{eqnarray}}
\def\eea{\end{eqnarray}}
\def\beas{\begin{eqnarray*}}
\def\eeas{\end{eqnarray*}}
\def\nn{\nonumber}
\def\ni{\noindent}
\def\G{\Gamma}
\def\d{\delta}
\def\l{\lambda}
\def\L{\Lambda}
\def\g{\gamma}
\def\m{\mu}
\def\n{\nu}
\def\s{\sigma}
\def\tt{\theta}
\def\b{\beta}
\def\a{\alpha}
\def\f{\phi}
\def\fh{\phi}
\def\y{\psi}
\def\z{\zeta}
\def\p{\pi}
\def\e{\epsilon}
\def\ve{\varepsilon}
\def\cl{{\cal L}}
\def\cv{{\cal V}}
\def\cz{{\cal Z}}
\def\co{{\cal O}}
\def\pl{\partial}
\def\ov{\over}
\def\~{\tilde}
\def\rar{\rightarrow}
\def\lar{\leftarrow}
\def\lrar{\leftrightarrow}
\def\rra{\longrightarrow}
\def\lla{\longleftarrow}
\def\8{\infty}
\def\h{\hbar}
\def\lg{\langle}
\def\rg{\rangle}
\def\agg{\biggl({m_\f^2 \ov \bar{\m}^2}\biggr)}
\def\ed{\end{document}}

\newcommand{\fr}{\frac}


\title{Reduction of a Class of Three-Loop Vacuum Diagrams to 
Tetrahedron Topologies}

\author{J.-M. Chung\footnote
  {Electronic address: jmchung@apctp.org}}
\address{Asia Pacific Center for Theoretical Physics, Seoul 130-012}
\author{B.~K. Chung\footnote
  {Electronic address: bkchung@khu.ac.kr}}
\address{Asia Pacific Center for Theoretical Physics, Seoul 130-012\\
and\\Research Institute for Basic Sciences
and Department of Physics,\\ 
Kyung Hee University, Seoul 130-701
{~}}

\maketitle
\draft
\begin{tighten}
\begin{abstract}

\indent
We obtain finite parts (as well as $\varepsilon$-pole parts) of massive 
three-loop vacuum diagrams with three-point and/or four-point interaction
vertices by reducing them to tetrahedron diagrams with both massive and 
massless lines, whose finite parts were given analytically in a recent 
paper by Broadhurst. 
In the procedure of reduction, the method of integration-by-parts 
recurrence relations is employed. We use our result to compute the 
$\overline{\rm MS}$ effective potential of the massive $\phi^4$ theory.

\end{abstract}
\end{tighten}

The calculations of the effective potential for a
single-component massive $\f^4$ theory \cite{cc1} and for a
massless $O(N)$ $\f^4$ theory \cite{jkps,cc2} were achieved at
the three-loop level in four-dimensional spacetime. (In three
dimensional spacetime, see the work of Rajantie \cite{rj} for 
a single-component $\f^4$ theory.)
The calculations in Ref.~1 and Refs.~2 and 3 are 
done in the dimensional regularization
scheme with a specific set of renormalization conditions. The
same calculations at a lower-loop level, in the cutoff
regularization, with the same renormalization conditions can be
found in Ref.~5 and Ref.~6 respectively. We see
that the results agree with each other. Therefore, in the
mass-dependent scheme, we do not need to calculate finite the 
parts of
three-loop diagrams. Knowledge of the pole terms is sufficient.

However, in a mass-independent scheme, such as the MS or
$\overline{\rm MS}$ scheme, we have to calculate three-loop
diagrams to the finite parts, i.e., to the $\ve^0$ order in the 
$\ve$ expansion. Without imposing renormalization
conditions at a specific scale, we just leave an arbitrary constant
$\m$, which is introduced inevitably for dimensional reasons, 
unspecified as in Eq.~(\ref{v0123}) below. This has the drawback 
that it does not involve true physical parameters measured at 
a given scale. Though it normally takes some effort to express 
physically measurable quantities in terms of the parameters of the
expression, the renormalization group (RG) equation is dealt with 
much easier,\footnote{See 
Ref.~7 for two-loop RG improvement of the effective potential.}
and the calculations in complicated theories are much more convenient.

The purpose of this note is to reduce a class of massive three-loop 
vacuum diagrams to (three-loop) tetrahedron diagrams with both massive 
and massless lines by using the method of recurrence relations. Once the 
reductions are completed, the finite parts (as well as the
$\varepsilon$-pole parts) of the diagrams in question can be readily 
obtained because the finite parts of these tetrahedron diagrams were
determined analytically by Broadhurst \cite{br2}. 

Let us define three-loop vacuum integrals $J$, $K$, and $L$,
which are nonfactorizable into lower-loop integrals:
 \bea
 J&\equiv&\int_{kpq}{1\ov (p^2+\s^2)[(p+k)^2+\s^2]
 (q^2+\s^2)[(q+k)^2+\s^2]}\;,\nn\\
 K&\equiv&\int_{kpq}{1\ov (k^2+\s^2)(p^2+\s^2)[(p+k)^2+\s^2]
 (q^2+\s^2)[(q+k)^2+\s^2]}\;,\nn\\
 L&\equiv&\int_{kpq}{1\ov (k^2+\s^2)^2(p^2+\s^2)[(p+k)^2+\s^2]
 (q^2+\s^2)[(q+k)^2+\s^2]}
\;.\label{dfn}
 \eea
The momenta in Eq.~(\ref{dfn}) are all (Wick-rotated) Euclidean,
and the abbreviated integration measure is defined as
 \bea
 \int_k=\m^{4-n}\int{d^n k\ov (2\p)^n}\;, \label{om}
 \eea
where $n=4-2\ve$ is the space-time dimension in the framework of
dimensional regularization, and $\m$ is an arbitrary constant with
a dimension of mass. The pole parts of all the integrals in Eq.~(\ref{dfn}) 
are known [1,~3]. Now, the problem is to the find the finite parts 
of these integrals in terms of the known transcendental numbers.

While massless multi-loop diagrams can be dealt with by
essentially algebraic methods \cite{ct}, the situation is more
complicated in the case of {\em massive} diagrams. In notation of
Avdeev \cite{av}, the above three-loop diagrams $J$, $K$, and $L$
correspond to $B_N(0,0,1,1,1,1)$, $D_5(1,1,1,1,1,0)$, and
$D_5(1,1,1,1,2,0)$, respectively. The method of integration-by-parts 
recurrence relations, given in Avdeev's paper \cite{av}, 
allows us to connect the integrals
$B_N(0,0,1,1,1,1)$, $D_5(1,1,1,1,1,0)$, and $D_5(1,1,1,1,2,0)$ to
the tetrahedron integrals $B_N(1,1,1,1,1,1)$, $B_M(1,1,1,1,1,1)$,
and $D_5(1,1,1,1,1,1)$. The analytic expressions of the finite
parts, i.e., $\ve^0$ order terms, for all tetrahedron vacuum
diagrams with different combinations of massless and massive
lines of a single mass scale were given in a recent paper by
Broadhurst \cite{br2}. With the convention of integration measure
used in Ref.~8,
 \beas \int [dk]=\int {d^n k\ov \s^{n-4}\p^{n/2}\G(1+\ve)},
 \eeas
we quote the results of Broadhurst which are relevant to our
calculation:
 \bea
 B_N(1,1,1,1,1,1)&=&V_{4N}={2\z(3)\ov \ve}+6\z(3)-14\z(4)-16U_{3,1}\;,\nn\\
 B_M(1,1,1,1,1,1)&=&V_{3T}={2\z(3)\ov \ve}+6\z(3)-9\z(4)\;,\nn\\
 D_5(1,1,1,1,1,1)&=&V_5={2\z(3)\ov \ve}+6\z(3)-{469\ov 27}
                    +{8\ov 3}{\rm Cl}_2^2\Bigl({\p\ov 3}\Bigr)
                    -16V_{3,1}
 \;, \label{bbdd}
 \eea
where $U_{3,1}$ and $V_{3,1}$ are defined as
 \beas
 &&U_{3,1}\equiv\sum_{m>n>0}{(-1)^{m+n}\ov m^3 n}\;,\nn\\
 &&V_{3,1}\equiv\sum_{m>n>0}{(-1)^m\cos(2\p n/3)\ov m^3 n}\;,
 \eeas
and can be expressed in terms of known transcendental
numbers as\footnote{The combination $V_{3,1}$ in Eq.~(\ref{uv})
in terms of the known transcendental numbers was found by Fleischer and
Kalmykov \cite{fk}.}
 \bea
 U_{3,1}&=&{\z(4)\ov 2}+{\z(2)\ov 2}\ln^2 2-{1\ov 12}\ln^4 2
 -2{\rm Li}_4\Bigl({1\ov 2}\Bigr)\;,\nn\\
 V_{3,1}&=&{1\ov 3}{\rm Cl}_2^2\Bigl({\p\ov 3}\Bigr)
 -{1\ov 4}\p{\rm Ls}_3\Bigl({2\p\ov 3}\Bigr)
 +{13\ov 24}\z(3)\ln 3-{259\ov 108}\z(4)+{3\ov 8}{\rm Ls}_4^{(1)}
 \Bigl({2\p\ov 3}\Bigr)\;. \label{uv}
 \eea
In the above equation, ${\rm Li}_4(x)$, ${\rm Cl}_2(x)$, ${\rm
Ls}_3(x)$, and ${\rm Ls}_4^{(1)}(x)$ are the polylogarithm,
Clausen's polylogarithm, the log-sine integral, and the generalized
log-sine integral, respectively \cite{lw}, whose numerical values
at the given arguments are
 \bea
 {\rm Li}_4\Bigl({1\ov 2}\Bigr)&=&0.517~479~061...\;,\nn\\
 {\rm Cl}_2\Bigl({\p\ov 3}\Bigr)&=&1.014~941~606...\;,\nn\\
 {\rm Ls}_3\Bigl({2\p\ov 3}\Bigr)&=&-2.144~762~212...\;,\nn\\
 {\rm Ls}_4^{(1)}\Bigl({2\p\ov 3}\Bigr)&=&-0.497~675~551...\;.
 \label{num}
 \eea

Using the the method of recurrence relations [9,~10,~13],
 we can arrive at following connections:
 \bea
 &&\s^{-4} B_N(0,0,1,1,1,1)=
   32B_4\biggl[{1\ov 2(1-3\ve)} + {4\ov 2-3\ve}
 -{2\ov 1-2\ve} - {1\ov 2(1-\ve)}\biggr]\nn\\
 &&~~~~- {486\ov 1-3\ve} - {729\ov 2(2-3\ve)}
 - {35\ov 2\ve} + {3\ov \ve^2} + {2\ov \ve^3}
 + {512\ov 1-2\ve} + {10\ov 1-\ve}\nn\\
 &&~~~~+ {\G(1-\ve)\G^2(1+2\ve)\G(1+3\ve)\ov \G^2(1+\ve)\G(1+4\ve)}
 \biggl[{14\ov 3\ve^2} + {35\ov \ve} + {378\ov 1-3\ve} + {189\ov 2-3\ve}
 - {896\ov 3(1-2\ve)} - {14\ov 3(1-\ve)} \biggr]\;,\nn\\
 &&D_5(1,1,1,1,1,1) = - \frac{2}{3} \s^{-2} D_5(1,1,1,1,1,0)
 \left[1 - \frac{1}{2 \ve}  \right] \nn \\
 &&~~~~-
  \frac{8}{3} B_4 \biggl[\frac{3}{2(1-3\ve)}
 -\frac{2}{1-2 \ve}  + \frac{1}{2(1-\ve)}\biggr]
 - \frac{2}{3\ve^2} \s^{-2} {\bf VL111}(1,1,1) \nn \\
 &&~~~~- \frac{2}{3 \ve^4} - \frac{1}{\ve^3}+ \frac{4}{3 \ve^2}
 + \frac{243}{2(1-3\ve)} + \frac{15}{\ve}
 - \frac{160}{3(1-2\ve)}+ \frac{7}{6(1-\ve)} \nn\\
 &&~~~~
 -\frac{\G(1-\ve)\G^2(1+2\ve)\G(1+3\ve)}{\G^2(1+\ve) \G(1+4\ve)}
 \biggl[ \frac{7}{9\ve^3}+\frac{14}{3\ve^2}
 + \frac{175}{9\ve}+\frac{189}{2(1-3\ve)}
 -\frac{224}{9(1-2\ve)}+ \frac{7}{18(1-\ve)}\biggr]\;,\nn\\
 &&D_5(1,1,1,1,2,0)={1\ov 3}\s^{-2}D_5(1,1,1,1,1,0)(1+\ve)
 -{32\ov 3}B_4\biggl[{1\ov 2(1-3\ve)}
 - {1\ov 1-2\ve} + {1\ov 2(1-\ve)}\biggr]\nn\\
 &&~~~~
 - {4\ov 3\ve}\s^{-2}{\bf VL111}(1,1,1) + {162\ov 1-3\ve}+ {44\ov 3\ve}
 - {4\ov 3\ve^3} - {256\ov 3(1-2\ve)} + {10\ov 3(1-\ve)}\nn\\
 &&~~~~
 - {\G(1-\ve)\G^2(1+2\ve)\G(1+3\ve)\ov \G^2(1+\ve)\G(1+4\ve)}
 \biggl[{28\ov 9\ve^2}+ {56\ov 3\ve} +{126\ov 1-3\ve}
 - {448\ov 9(1-2\ve)} + {14\ov 9(1-\ve)}\biggr]\;, \label{rec}
 \eea
where $B_4$ is proportional to
the difference between $B_N(1,1,1,1,1,1)$ and
$B_M(1,1,1,1,1,1)$ \cite{br1}, i.e.,
 \beas
 B_4 &=&-{(1-2\ve)(2-2\ve)\ov 4}\Bigl[B_M(1,1,1,1,1,1)
 -B_N(1,1,1,1,1,1)\Bigr]\;.
 \eeas
In Eq.~(\ref{rec}), {\bf VL111}(1,1,1) denotes a two-loop bubble
integral, shown in Fig.~1 of the paper by Fleischer and Kalmykov
\cite{fk}. Its value up to $\ve^3$ order in the $\ve$
expansion can be found in Eq.~(7) of Ref.~11. For all
higher-order terms, one may refer to Ref.~14.
For our desired accuracy, it is sufficient to take its value up to 
$\ve$ order:
 \bea
 \s^{-2}{\bf VL111}(1,1,1)&=&-{3\ov 2(1-\ve)(1-2\ve)}\biggl[
 {1\ov \ve^2}-{4\ov \sqrt{3}}{\rm Cl}_2\Bigl({\p\ov 3}\Bigr)\nn\\
 &&+\ve\biggl\{{4\ov \sqrt{3}}{\rm Cl}_2\Bigl({\p\ov 3}\Bigr)\ln 3
 -{\p^3\ov 3\sqrt{3}}-2\sqrt{3}{\rm Ls}_3\Bigl({2\p\ov 3}\Bigr)\biggr\}
 +O(\ve^2)\biggr]\;. \label{vl111}
 \eea

From Eqs.~(\ref{bbdd}) --- (\ref{vl111}), we eventually obtain
 \bea
 B_N(0,0,1,1,1,1)&=&{2\ov \ve^3}+{23\ov 3\ve^2}+{35\ov 2\ve}
 +{275\ov 12}+O(\ve)\;,\nn\\
 D_5(1,1,1,1,1,0)&=&-{1\ov \ve^3}-{17\ov 3\ve^2}
 +{1\ov \ve}\biggl[-{67\ov 3}+{12\ov \sqrt{3}}
 {\rm Cl}_2\Bigl({\p\ov 3}\Bigr)\biggr]-{229\ov 3}+{60\ov \sqrt{3}}
 {\rm Cl}_2\Bigl({\p\ov 3}\Bigr)\nn\\
 &&-{12\ov \sqrt{3}}{\rm Cl}_2\Bigl({\p\ov 3}\Bigr)\ln 3
 +{\p^3\ov \sqrt{3}}+6\z(3)
 +6\sqrt{3}{\rm Ls}_3\Bigl({2\p\ov 3}\Bigr)+O(\ve)\;,\nn\\
 D_5(1,1,1,1,2,0)&=&{1\ov 3\ve^3}+{2\ov 3\ve^2}
 +{1\ov \ve}\biggl[{2\ov 3}
 -{4\ov \sqrt{3}}{\rm Cl}_2\Bigl({\p\ov 3}\Bigr)\biggr]\nn\\
 &&-{2\ov 3}+{4\ov \sqrt{3}}{\rm Cl}_2\Bigl({\p\ov 3}\Bigr)\ln 3+2\z(3)
 -{\p^3\ov 3\sqrt{3}}-2\sqrt{3}{\rm Ls}_3\Bigl({2\p\ov 3}\Bigr)
 +O(\ve)\;.
 \eea
We readily recover the values of $J$, $K$, and $L$
in our original integration  measure, Eq.~(\ref{om}), by using the following
relations:
 \beas
 J&=&{\s^4\ov (4\p)^6}\biggl({\s^2\ov 4\p\m^2}\biggr)^{\!\!\!{-3\ve}}
 \exp\biggl[-3\g\ve
 +{3\z(2)\ov 2}\ve^2-\z(3)\ve^3+\cdots\biggr]B_N(0,0,1,1,1,1)\;,\nn\\
 K&=&{\s^2\ov (4\p)^6}\biggl({\s^2\ov 4\p\m^2}\biggr)^{\!\!\!{-3\ve}}
 \exp\biggl[-3\g\ve
 +{3\z(2)\ov 2}\ve^2-\z(3)\ve^3+\cdots\biggr]D_5(1,1,1,1,1,0)\;,\nn\\
 L&=&{1\ov (4\p)^6}\biggl({\s^2\ov 4\p\m^2}\biggr)^{\!\!\!{-3\ve}}
 \exp\biggl[-3\g\ve
 +{3\z(2)\ov 2}\ve^2-\z(3)\ve^3+\cdots\biggr]D_5(1,1,1,1,2,0)
 \;,
 \eeas
where the $\exp[~]$ factor comes from an expansion of $\G^3(1+\ve)$
and $\g$ is the Euler constant.

In the standard $\overline{\rm MS}$ scheme \cite{muta}, the
factors $\ln(4\p)$ and $\g$ are absorbed into the
renormalization scale $\m$. Howerever, in the other widespread 
$\overline{\rm MS}$ convention (see, e.g., Refs.~16 and 17), the factor
$\z(2)$ is absorbed further into the scale $\m$. (This convention
gives the same result in the one-loop diagrams and is more
convenient in higher-loop massive calculations.) By introducing a
new renormalization scale $\bar{\m}$,
 \bea
 \bar{\m}^2=4\p\m^2\exp\biggl[-\g+{\z(2)\ve\ov 2}\biggr]\;,\label{mms}
 \eea
the above three integrals $J$, $K$, and $L$ are given as
follows:
 \bea
 J&=&{\s^4\ov (4\p)^6}\biggl({\s^2\ov \bar{\m}^2}
 \biggr)^{\!\!\!{-3\ve}}\biggl[{2\ov \ve^3}+{23\ov 3 \ve^2}
 +{35\ov 2\ve}+F_J\biggr]\;,\nn\\
 K&=&{\s^2\ov (4\p)^6}\biggl({\s^2\ov \bar{\m}^2}
 \biggr)^{\!\!\!{-3\ve}}\biggl[-{1\ov \e^3}-{17\ov 3\ve^2}
 +{1\ov \ve}\biggl\{-{67\ov 3}+{12\ov \sqrt{3}}
 {\rm Cl}_2\Bigl({\p\ov 3}\Bigr)\biggr\}
 +F_K\biggr]\;,\nn\\
 L&=&{1\ov (4\p)^6}\biggl({\s^2\ov \bar{\m}^2}
 \biggr)^{\!\!\!{-3\ve}}\biggl[{1\ov 3\ve^3}+{2\ov 3 \ve^2}
 +{1\ov \ve}\biggl\{{2\ov 3}
 -{4\ov \sqrt{3}}{\rm Cl}_2\Bigl({\p\ov 3}\Bigr)\biggr\}
 +F_L\biggr]\;,\label{jklm}
 \eea
where
 \bea
 &&F_J={275\ov 12}-2\z(3)\;,\nn\\
 &&F_K=-{229\ov 3}+{\p^3\ov \sqrt{3}}
 +7\z(3)+{60\ov \sqrt{3}}{\rm Cl}_2\Bigl({\p\ov 3}\Bigr)
 -{12\ov \sqrt{3}}{\rm Cl}_2\Bigl({\p\ov 3}\Bigr)\ln 3
 +6\sqrt{3}{\rm Ls}_3\Bigl({2\p\ov3}\Bigr)\;,\nn\\
 &&F_L=-{2\ov 3}-{\p^3\ov 3\sqrt{3}}+{5\z(3)\ov 3}+{4\ov \sqrt{3}}
 {\rm Cl}_2\Bigl({\p\ov 3}\Bigr)\ln 3
 -2\sqrt{3}{\rm Ls}_3\Bigl({2\p\ov3}\Bigr)
 \;.\label{fjklm}
 \eea
This completes the analytic evaluations of the three-loop vacuum
integrals of Eq.~(\ref{dfn}) up to the finite parts.

The purely numerical computation of the finite parts for some three-loop
vacuum diagrams in a paper by Pelissetto and Vicari \cite{pv}
enables us to extract the numerical values for $F_J$, $F_K$,
and $F_L$. In obtaining $F_J$ and $F_K$, we assume first
the {\em unknown} finite parts $F_J$ and $F_K$ for $J$ and $K$ in
Eq.~(\ref{jklm}) since the pole parts are already known
\cite{cc1,cc2}. Then, we differentiate $J$ and $K$, twice and once, 
respectively, with respect to $\s^2$. Meanwhile, we can differentiate 
$J$ and $K$ in Eq.~(\ref{dfn}), twice and once, respectively, 
with respect to $\s^2$ before the momentum integrations, 
yielding the three-loop diagrams given in Appendix B of
Ref.~18 whose finite parts have been numerically calculated.
By equating the results of the differentiations thus
done, we determine the unknown values of $F_J$ and $F_K$.
The extracted results are
 \bea
 &&F_J={275\ov 12}-6\sqrt{3}{\rm Cl}_2\Bigl({\p\ov 3}\Bigr)-22\z(3)+80S_1
 +20S_2-120S_5-30S_6-6S_7\;,\nn\\
 &&F_K=-{229\ov 3}+{28\ov \sqrt{3}}{\rm Cl}_2\Bigl({\p\ov 3}\Bigr)
 -\p^2+9\z(3)-24S_1-6S_2+6S_4-12S_7\;,\nn\\
 &&F_L=-{2\ov 3}-4\sqrt{3}{\rm Cl}_2\Bigl({\p\ov 3}\Bigr)-{\p^2\ov 3}
 -{5\z(3)\ov 3}+8S_1+2S_2+2S_4
 \;, \label{nm}
 \eea
where the quantities $S_1$, $S_2$, $S_4$, $S_5$, $S_6$, and $S_7$
were calculated numerically in Appendix B of Ref.~18.
We see that the numerical values of $F_J$, $F_K$, and $F_L$  given in
Eq.~(\ref{nm}) agree with the analytical values
in Eq.~(\ref{fjklm}).\footnote{Also, in a recent paper by Anderson et al. 
\cite{abs}, $F_J$ was calculated numerically. The numerical value of
$C_0$ in Eq.~(18) of Ref.~19 agrees with our analytic value $F_J$
in Eq.~(\ref{fjklm}): our $F_J+23\p^2/12$ means $C_0$ of Ref.~19.}

Our results, Eqs.~(\ref{jklm}) and (\ref{fjklm}), together with the result 
for the all-massive-line tetrahedron diagram in Ref.~8, 
\beas
 M&\equiv&\int_{kpq}{1\ov (k^2+\s^2)(p^2+\s^2)(q^2+\s^2)
 [(k-p)^2+\s^2][(p-q)^2+\s^2][(q-k)^2+\s^2]}\nn\\
 &=&{1\ov (4\p)^6}\biggl({\s^2\ov \bar{\m}^2}
 \biggr)^{\!\!\!{-3\ve}}\biggl[{2\z(3)\ov \ve}
 +6\z(3)-17\z(4)-{2\p^2\ov 3}\ln^2 2+{2\ov 3}\ln^4 2
 -4{\rm Cl}_2^2\Bigl({\p\ov 3}\Bigr)
 +16{\rm Li}_4\Bigl({1\ov 2}\Bigr)\biggr]\;,\nn\\
\eeas
enable us to calculate the three-loop effective potential 
in the $\overline{\rm MS}$ scheme for the single-component massive 
$\f^4$ theory. The renormalization of the three-loop effective potential 
is straightforward, albeit long. Thus, we simply report the result:
 \bea
 V&=&V^{(0)}+\h V^{(1)}+\h^2 V^{(2)}+\h^3 V^{(3)}+O(\h^4)\;,\nn\\
 V^{(0)}&=&{m^2 \f^2\ov 2}+{\l \f^4\ov 4!}\;,\nn\\
 V^{(1)}&=&{\l\ov (4\p)^2}\biggl[-{3m^4\ov 8\l}-{3 m^2 \f^2\ov 8}
 -{3\l\f^4\ov 32}
 +\biggl\{{m^4\ov 4\l}+{m^2 \f^2\ov 4}
 +{\l\f^4\ov 16}\biggr\}\ln\agg\biggr]\;,\nn\\
 V^{(2)}&=&{\l^2\ov (4\p)^4}\biggl[{m^4\ov 8\l}
 +m^2\f^2\biggl( {3\ov 4}-{1\ov 2\sqrt{3}}
 {\rm Cl}_2\Bigl({\p\ov 3}\Bigr)\biggr)
 +\l\f^4\biggl({11\ov 32}-{1\ov 4\sqrt{3}}
 {\rm Cl}_2\Bigl({\p\ov 3}\Bigr)\biggr)\nn\\
 &&-\biggl\{{m^4\ov 4\l}+{3m^2\f^2\ov 4}
 +{5\l\f^4\ov 16}\biggr\}\ln\agg
 +\biggl\{{m^4\ov 8\l}+
 {m^2 \f^2\ov 4}+{3\l\f^4\ov 32}\biggr\}\ln^2\agg\biggr]\;,\nn\\
 V^{(3)}&=&{\l^3\ov (4\p)^6} \biggl[{1\ov 576}{m^4\ov \l}
 +m^2\f^2\biggl(-{2363\ov 576}
 +{13\ov 4\sqrt{3}}{\rm Cl}_2\Bigl({\p\ov 3}\Bigr)
 +{3\z(3)\ov 4}\biggr)\nn\\
 &&+\l\f^4\biggl(-{4487\ov 2304}
 +{11\ov 8\sqrt{3}}{\rm Cl}_2\Bigl({\p\ov 3}\Bigr)
 + {1\ov 6}{\rm Cl}_2^2\Bigl({\p\ov 3}\Bigr)
 -{2\ov 3}{\rm Li}_4\Bigl({1\ov 2}\Bigr) +{17\z(4)\ov 24}\nn\\
 &&+{\p^2\ln^2 2\ov 36}-{\ln^4 2\ov 36}\biggr)
 +\biggl\{{41m^4\ov 96\l}
 +m^2\f^2\biggl( {371\ov 96}-{7\ov 4\sqrt{3}}
 {\rm Cl}_2\Bigl({\p\ov 3}\Bigr)\biggr)\nn\\
 &&+\l\f^4\biggl({701\ov 384}
 -{3\sqrt{3}\ov 4}{\rm Cl}_2\Bigl({\p\ov 3}\Bigr)
 +{\z(3)\ov 4}\biggr)\biggr\}\ln\agg
 -\biggl\{{17m^4\ov 48\l}+{37m^2\f^2\ov 24}\nn\\
 &&+{143\l\f^4\ov 192}\biggr\}\ln^2\agg
 +\biggl\{{5m^4\ov 48\l}+{7m^2\f^2\ov 24}
 +{9\l\f^4\ov 64}\biggr\}\ln^3\agg\biggr]\;, \label{v0123}
 \eea
where $m_\f^2$ is defined as $m_\f^2\equiv m^2+{\l\f^2\ov 2}$.

In summary, using the method of integration-by-parts recurrence 
relations, we have obtained the exact relations between 
the non-tetrahedron three-loop integrals $B_N(0,0,1,1,1,1)$, 
$D_5(1,1,1,1,1,0)$, and $D_5(1,1,1,1,2,0)$ and the tetrahedron 
three-loop integrals $B_N(1,1,1,1,1,1)$, $B_M(1,1,1,1,1,1)$, and 
$D_5(1,1,1,1,1,1)$, whose values are known to the finite parts.  
As an application of our loop calculations, the analytic evaluation 
of three-loop effective potential in the $\overline{\rm MS}$ scheme for 
the single-component massive $\f^4$ theory was obtained.

\begin{center}
{\bf ACKNOWLEDGEMENT}
\end{center}

This work was supported by a Korea Research Foundation grant 
(KRF-2000-015-DP0066).
\begin{tighten}

\end{tighten}
\end{document}